\font\subtit=cmr12
\font\name=cmr8
\input harvmac

\def\plb#1#2#3#4{#1, {\it Phys. Lett.} {\bf {#2}}B (#3), #4}
\def\npb#1#2#3#4{#1, {\it Nucl. Phys.} {\bf B{#2}} (#3), #4}
\def\npbib#1#2#3{{\it Nucl. Phys.} {\bf B{#1}} (#2), #3}
\def\prl#1#2#3#4{#1, {\it Phys. Rev. Lett.} {\bf {#2}} (#3), #4}

\def\cmp#1#2#3#4{#1, {\it Comm. Math. Phys.} {\bf {#2}} (#3), #4}

\def\mpl#1#2#3#4{#1, {\it Mod. Phys. Lett.} {\bf A{#2}} (#3), #4}
\def\ijmpa#1#2#3#4{#1, {\it Int. Jour. Mod. Phys.} {\bf A{#2}} (#3), #4}
\def\jmp#1#2#3#4{#1, {\it Jour. Math. Phys.} {\bf {#2}} (#3), #4}

\def\jgp#1#2#3#4{#1, {\it Journal Geom. Phys.} {\bf {#2}} (#3), #4}

\null\vskip 3truecm
\centerline{\subtit
BOSONIC FIELD PROPAGATORS ON ALGEBRAIC CURVES
}
\vskip 1truecm
\centerline{F{\name RANCO} F{\name ERRARI}$^{a}$ and J{\name AN}
S{\name OBCZYK}$^b$}
\smallskip $^a${\it Dipartimento di Fisica,
Universit\`a di Trento, I-38050 Povo (TN), Italy\foot{E-mail:
fferrari@science.unitn.it}.
}
\smallskip $^b${\it
The Abdus Salam International Centre for Theoretical Physics,
P.O.Box 586, 34-100 Trieste, Italy\foot{On leave on absence from
Institute for Theoretical Physics, Wroc\l aw
University, pl.  Maxa Borna 9, 50204 Wroc\l aw, Poland, E-mail:
jsobczyk@proton.ift.uni.wroc.pl}.}
\smallskip
\vskip 3cm
\centerline{ABSTRACT}
{\narrower \abstractfont
In this paper we investigate massless scalar field theory on non-degenerate
algebraic
curves.
The propagator is written
in terms
of the parameters appearing in the polynomial defining
the curve.
This provides an alternative to the language of
theta functions.
The main result
is a derivation of the third
kind differential normalized in such a way that its periods around the homology cycles
are purely imaginary.
All the physical correlation functions of
the scalar fields can be expressed in terms of
this object.
This paper contains a detailed analysis of the techniques
necessary to study field theories on algebraic curves.
A simple expression
of the scalar field propagator is found in a
particular case in which the algebraic curves
have $Z_n$ internal symmetry and one of the fields is located at a
branch point.
} 
\Date{September 1999}
\pageno=0
\vfill\eject
\newsec {INTRODUCTION}
\vskip 1cm

In the last decade there has been a growing interest in the applications
of algebraic curves to several problems in theoretical physics,
ranging from the theory of strings to condensed matter physics
\ref\hyel{
\npb{D.  Lebedev and A.  Morozov}{302}{1986}{163}; \plb
{E. Gava, R. Iengo and G. Sotkov}{207}{1988}{283}; \plb{A. Yu. Morozov and
A. Perelomov}{197}{1987} {115};
F. Ferrari, {\it Fizika} {\bf 21} (1989), 32; \mpl{J. Sobczyk}{6}{1991}{1103};
\npb{D.  Montano}{297}{1988}{125};
\ijmpa{M.
A.  Bershadsky and A.  O.  Radul}{2}{1987}{165}.}\nref\zam{
\npb{Al. B. Zamolodchikov}{285}{1987}{481}.}\nref\iezhu{\npb{E. Gava,
R.  Iengo and C.  J.  Zhu}{323}{1989}{585};
\plb{R.  Jengo and C.  J.  Zhu}{212}{1988}{313}.}\nref\knirev{V. G. Knizhnik,
{\it Sov. Phys. Usp.} {\bf 32}(11)
(1989) 945.}\nref\knione{\cmp{V. G. Knizhnik}
{112}{1987} {587}.}\nref\bhs{L. Borisov, M.B. Halpern and C. Schweigert,
{\it Int. Jour. Mod. Phys.} {\bf A13} (1)
(1998), 125.}\nref\apieft{S. A. Apikian and C. J. Efthimiou, {\it Int. Jour.
Mod. Phys.}, {\bf A12} 1997, 4291, hep-th/9610051.}\nref\lerche{W. Lerche,
S. Stieberger
and N. P. Warner, {\it Quartic Gauge Couplings from K3 Geometry},
Preprint CERN-TH/98-378, hep-th/9811228;
W. Lerche and S. Stieberger, {\it Adv. Theor. Math. Phys.} {\bf 2} (1998),
1105, hep-th/9804176.}\nref\iezhn{R. Iengo and C.-J. Zhu,
{\it Explicit Modular Invariant Two-Loop Superstring Amplitude Relevant
for $R^4$}, JHEP {\bf 9906} (1999), 011.}\nref\kholvil{A. L. Kholodenko
and T. A. Vilgis, {\it Phys. Rep.} {\bf 298} (1998), 251;
A. L. Kholodenko, {\it Random Walks on Figure Eight: from Polymers through
Chaos to Gravity and Beyond}, cond-mat/9905221.}--\ref\nechaev{S. Nechaev,
{\it Int. Jour. Mod. Phys.} {\bf B4} (1990), 1809;
{\it Statistics of Knots and Entangled Random Walks},
extended version of lectures presented at Les Houches
1998 Summer School on {\it Topological Aspects of Low
Dimensional Systems}, July 7 - 31, 1998, cond-mat/9812205.}.
Algebraic curves, defined as the region in which a
complex polynomial of two variables vanishes,
provide an explicit and useful representations
of Riemann surfaces.
Thus, one can take advantage of well established
results and theorems of algebraic geometry
\ref\fakra{H. Farkas and I. Kra, {\it Riemann Surfaces},
Springer Verlag, 1980.}\nref\fay{J.  D.  Fay, Theta Functions on Riemann
Surfaces, Lecture Notes in Mathematical Physics no.  352, Springer Verlag,
1973.}--\ref\grha{
P. Griffiths
and J. Harris, {\it Principles of Algebraic
Geometry}, John Wiley \& Sons, New York 1978.}.
So far, mainly applications involving
curves with cyclic monodromy group have been studied
neglecting other, more interesting
examples of algebraic curves.
The reason is that computations on general
algebraic curves are intrinsically
complicated.
For instance, it is not known how to express
the theta functions \fakra\
in terms of the parameters
appearing in the polynomial which define the curve. This is a
great handicap, since theta functions
are the main building blocks in the construction
of tensors with given zeros and poles on Riemann surfaces.
Other problems are connected with the analytical continuation
of multivalued functions on the complex plane.
For example,
any attempt of setting up an operator formalism on algebraic curves
based on the
monodromy of the fields \knirev,\ref\brrmp{
M. A. Bershadsky and A. Radul,
{\it Phys. Lett.} {\bf 193B} (1987), 21.},
leads to differential equations of the
Riemann monodromy problem type which are too difficult to
be solved in practice.
On the contrary,
there exist several ways
to expand free conformal
fields on a Riemann surface using the formalism of theta
functions \ref\kn{I. M.
Krichever and S. P. Novikov, {\it Funk. Anal. Pril.} {\bf 21} No.2 (1987), 46;
{\bf 21} No.4 (1988), 47.}\nref\blmr{L. Bonora, A. Lugo,
M. Matone and J. Russo,
{\it Comm. Math. Phys.} {\bf 123} (1989), 329.}\nref\bontwo{\plb{L. Bonora, M.
Matone, F. Toppan and K. Wu}{224}{1989}{115}; \npbib{334}{1990}{717}; L. Bonora
and F. Toppan, {\it Rev. Math. Phys.} {\bf 4} (1992), 429.}\nref\agr{E.
Date, M. Jimbo, M. Kashiwara and T. Miwa, In: Proc.
of International Symposium on Nonlinear Integrable Systems, Kyoto 1981,
M. Jimbo and T. Miwa (eds.), Singapore (1983); \prl{S. Saito}{36}{1987}{1819};
L. Alvarez-Gaum\'e, C. Gomez and C. Reina, New
Methods in String Theory, in: Superstrings '87, L. Alvarez-Gaum\'e
(ed.), Singapore, World Scientific 1988; \mpl{N. Ishibashi, Y. Matsuo and
Y. Ooguri}{2}{1987}{119}; N. Kawamoto, Y. Namikawa, A.
Tsuchiya and Y. Yamada, {\it Comm. Math. Phys.} {\bf 116} (1988),
247.}\nref\cvafa{\plb{C. Vafa}{190}{1987}{47}.}\nref\raina{A. K. Raina,
{\it Comm. Math. Phys.} {\bf 122} (1989), 625; {\it ibid.} {\bf 140}
(1991), 373; {\it Lett. Math. Phys.} {\bf 19} (1990), 1;
{\it Expositiones Mathematicae} {\bf 8} (1990), 227; {\it Helvetica
Physica Acta} {\bf 63} (1990), 694.}\nref\pdvpz{
P. di Vecchia,
{\it Phys. Lett.} {\bf B248} (1990), 329; P. Di Vecchia, F. Pezzella,
M. Frau, K. Hornfleck, A. Lerda and A. Sciuto, {\it Nucl. Phys.}
{\bf B332} (1989), 317; {\it ibid.} {\bf B333} (1990), 635; A. Clarizia
and F. Pezzella, {\it Nucl. Phys.} {\bf B298} (1988), 636;
\plb{G. Cristofano, R. Musto, F. Nicodemi and R. Pettorino}{217}{1989}
{59}.}\nref\russo{\npb{A. Lugo and J. Russo}{322}{1989}{210};
\plb{J.Russo}{220}{1989}{104}.}--\ref\semi{A. M.
Semikhatov, {\it Phys. Lett.} {\bf B212} (1988), 357;
O. Lechtenfeld, {\it Phys.
Lett} {\bf B232} (1989) 193; \npb{U. Carow-Watamura and S. Watamura}{288}
{1987}{500}; \npbib{301}{1988}{132}; \npbib{302}{1988}{149}; \npbib{308}{1988}
{143}; U. Carow-Watamura, Z. F. Ezawa, K.
Harada, A. Tezuka and S. Watamura, {\it Phys. Lett.} {\it B227}
(1989), 73.}.

Some progress in the physical applications of general algebraic curves can
be obtained by studying simple conformal field theories.
For example, the correlation functions of the $b-c$ systems have been
explicitly
computed by means
of an operator formalism in
\ref\naszejdwa{
\ijmpa{F. Ferrari and J. Sobczyk}{11}{1996}{2213}.}--\ref\naszed{
\jgp{F. Ferrari and J. Sobczyk}{19}{1996}{287}.}.
The construction of a generalized Laurent basis in order to expand meromorphic
tensors has been achieved in \naszed.
In the particular case of nonabelian monodromy groups $D_n$,
$n=2,3,4,\ldots$, it is possible
to show that these elements are connected to a Riemann monodromy problem
\ref\fercmp{\cmp{F. Ferrari}{156}{1993}{179}.}.
The relations between free $b-c$ systems on algebraic curves and new conformal
field theories on the complex plane has been explored in
\ref\naszejone{\jmp{F.  Ferrari, J.  Sobczyk and W.  Urbanik}
{36}{1995}{3216}, hep-th/9310102.},\ref\naszejgp{\jgp{F. Ferrari and
J. Sobczyk}{29}{1999}{161}, hep-th/9709162.}.
The operator formalism of \naszed\ has been extended to the $\beta-\gamma$
systems \ref\fms{
\npb{D. Friedan, E. Martinec and S.
Shenker}{271}{1986}{93}.},\ref\ag{
\npb{L. Alvarez-Gaum\' e, C. Gomez, P. Nelson, G. Sierra and C. Vafa}
{311}{1988}{333}.} with integer spins in ref.
\ref\ferjanjmp{F. Ferrari and J. T. Sobczyk,
{\it Operator Formalism for Bosonic
Beta--Gamma Fields on General Algebraic Curves},
{\it Jour. Math. Phys.} {\bf 39} (10) (1998), 5148.}.

In this paper we study correlation functions
of free massless scalar field theory on algebraic curves.
The form of the curve is restricted only by a requirement of non-degeneracy.
The problem is reduced to the computation of a differential
of the third kind normalized so that its periods along
the homology cycles are purely imaginary \fay.
To this purpose, we construct here a differential of the third kind
using the techniques developed in ref. \naszed.
Its periods along the homology cycles are then fixed by adding
suitable linear combinations of holomorphic differentials.
To pick up exactly the linear combination
which makes all the periods imaginary is however not a simple task.
As a matter of fact, 
on general algebraic curves it is not even known how to construct a basis
of homology cycles.
To overcome this difficulty, we exploit here a set of Riemann bilinear
identities
\fakra, the derivation of which is presented for a sake of completness 
in Appendix A. These relations uniquely determine
the desired differential of the third kind and provide an explicit
expression for it in terms of the parameters which characterize
the curve\foot{Let us notice that similar methods, using integral
identities like the Riemann bilinear relations, have been already
used in the computation of scalar
Green functions on hyperelliptic curves \ref\iezhn,\ref\knirieimp{V.
G. Knizhnik,
{\it Phys. Lett.} {\bf 196B} (1987), 473.}.}.

The material contained in this paper is organized as follows.
In Section 2 we develop the necessary techniques
to study field theories on algebraic curves.
With respect to ref. \naszed, the curves are entirely general
apart from the
non-degeneracy. To simplify our analysis, the defining polynomials
are reduced to a canonical form introducing a set
of suitable transformations. The relevant differentials like the
Weierstrass kernel and the holomorphic differentials are derived together
with their divisors. The genus of the curve is obtained
from the number of holomorphic differentials. This result is
checked using a Baker's method 
\ref\forsyth{A. R. Forsyth, {\it Theory of Functions of a Complex
Variable}, Vols I and II, Dover Publications, Inc., New York, 1965.}.
In Section 3 the theory of massless scalar field theories is briefly
reviewed following
\ref\bonien{M. Bonini and R. Jengo, {\it Int. Jour. Mod. Phys.} {\bf A3}
(1988), 841.}.
It is shown that all the correlation functions can be written in terms of
the canonical differential of the
third kind with
purely imaginary periods around the non-trivial
monodromy cycles.
This canonical differential is constructed as a
a linear combination of Weierstrass kernels and holomorphic differentials.
The coefficients of the linear combination
are determined by means of a system
of Riemann bilinear equations, which are satisfied if and only if the
differential of the third kind has purely imaginary periods.
The correlation functions of the scalar fields
obtained in this way explicitly depend on the
parameters of the
polynomial which defines the algebraic curve as desired.
Some examples are worked out in Section 4.
The latter Section contains also the derivation of the massless scalar field
propagator 
on $Z_n$ symmetric algebraic curves \knirev\ supposing that one of the scalar
fields is located at a branch point.
This particular case is of some relevance in string theories
as explained in ref. \lerche.
The Riemann bilinear relations used in Section 3 are derived in Appendix
A, while in Appendix B it is shown how integrals on an algebraic curve
can be rewritten as integrals of multivalued volume forms
over the complex sphere following \grha.
Finally, we discuss our results and their possible applications.

\newsec {THE LANGUAGE OF ALGEBRAIC CURVES}

Let {\bf C} be the
complex plane and {\bf CP}$^1$ the complex projective line
which coincides
with the compactified complex
plane {\bf C}$\cup\{\infty\}$.
We will describe Riemann surfaces as $n-$sheeted branched covers of
${\bf CP}^1$ \grha. The latter are given as
algebraic curves defined as the locus
of points $(z,y)\in {\bf CP}^1\times {\bf CP}^1$ for which the
following equation is satisfied:
\eqn\curve{F(z,y)=0}
Here $F(z,y)$ denotes a polynomial
of the form:
\eqn\weipd{F(z,y) = P_n(z)y^n + P_{n-1}(z)y^{n-1} +
\ldots + P_1(z)y + P_0(z) = 0}
with
$P_s(z)=\sum\limits_{m=0}^{n_s}A_{s,m}z^m$ for $s=0,\ldots,n$
and $n_s\in{\bf N}$.
By a well known theorem, apart from subtleties coming from singular points,
any Riemann surface can be expressed
as an algebraic curve of this kind.
Thus, from now on, we will use the words Riemann surface, 
$n-$sheeted branched covers of
${\bf CP}^1$ and algebraic curves interchangeably.
The best known algebraic curves are the
hyperelliptic curves, whose polynomials $F(z,y)$ are simply given by:
$F(z,y) =y^2-P_0(z)$.
Also the slightly more general $Z_n$ {symmetric
curves} will be often mentioned here.
They are characterized by
$F(z,y)=y^n-P_0(z)$.

One can solve eq. \curve\ with respect to $y$ and obtain in this way a
function $y(z)$ which is multivalued on ${\bf CP}^1$. Its
$n$ branches are interchanged at a set of $N_{bp}$ branch points
$a_i,\ldots,a_{N_{np}}\in {\bf CP}^1$ as will be discussed  below.
In the following, the branches of $y(z)$ will be denoted with the
symbol $y^{(\alpha)}(z)$, $\alpha=0,\ldots,n-1$ and the
first letters of the Greek alphabet will be used as branch indices.
To simplify our analysis, we assume that $P_n(z)=1$ in eq. \weipd.
There is no loss of generality in this assumption.
As a matter of fact, if $P_n(z)\ne 1$, one can always perform the change of
variables
\eqn\firtra{\tilde y(z)=y(z)P_n(z)}
which does not affect the monodromy properties of $y(z)$, so that both
$\tilde y(z)$ and $y(z)$ are meromorphic functions on the same Riemann surface.
However,
it is easy to realize that $\tilde y$ satisfies
the equation
$\sum_{i=0}^n\tilde P_{n-i}(z) \tilde y^{n-i}=0$,
where now $\tilde P_n(z)=1$ as desired and $\tilde
P_{n-i}(z)=P_{n-i}(z)P_n^{i-1}(z)$ for
$i=1,\ldots,n$.

Further, we require that none of the branch points is in $z=\infty$.
Again, this condition does not restrict the generality of our discussion as
we will show below.
First of all, the presence of a branch point at infinity can be detected
investigating the fact that, for large values of $z$, the function $y(z)$
exhibits the following behavior:
 \eqn\ansatz{y(z)_{\sim\atop z\rightarrow\infty}c z^p+{\rm lower}\enskip
{\rm  order} \enskip {\rm terms}}
Substituting eq. \ansatz\
in eq. \curve\ and solving the latter at the leading
order in $z$, one determines the allowed values of the constants $c$ and $p$.
A branch point at $z=\infty$ is indicated by noninteger solutions
for $p$.
If this is the case, it is always possible to perform a
$SL(2,{\bf C})$ transformation in $z$, in such a way that
the branch point at infinity is moved to a finite region of the plane.
Of course, the condition $P_n(z)=1$ does no longer hold in the
new variable, but it can be easily restored with the
aid of the transformations \firtra\ in $y$.
Since the letter transformation does not affect the
monodromy properties of $y$, the branch point at $z=\infty$ cannot be
reintroduced.

We are now ready to study the finite branch points $a_1,\ldots,a_{N_{bp}}$.
Supposing that the curve \curve\ is nondegenerate\foot{See for instance
ref. \grha\
for the definition of nondegeneracy.}, they are the solutions of
of the following system of equations:
\eqn\bpdef{F(z,y)=F_y(z,y)=0}
where $F_y(z,y)=dF(z,y)/dy$.
It is useful to eliminate from eq. \bpdef\
the variable $y$.
As an upshot, one obtains a polynomial equation in $z$
of the kind $r(z)=0$. Apart from very special curves,
in which $r(z)$ has multiple roots,
its degree coincides with the number of branch points $N_{bp}$.
Let us notice that it is possible to derive the resultant $r(z)$
of eqs. \bpdef\ explicitly using the dialitic
method of Sylvester
\ref\cenr{F.  Enriques and O.
Chisini, {\it Lezioni sulla Teoria
Geometrica delle Equazioni e delle Funzioni Algebriche}, Zanichelli,
Bologna (in italian).}.

To each branch point $a_i$ one can associate an integer $\nu_i$, called
the ramification index and defined as the number of branches of $y(z)$
that are exchanged at that branch point.
Clearly, $2\le\nu_i\le n$.
At a branch point of
ramification index $\nu_i$,  the polynomial $F(z,y)$
vanishes together with its first $\nu_i-1$ partial derivatives
in $y$. 
The genus $g$ of the Riemann surface \curve\ is related to the
ramification indices
of the branch points and the number of sheets $n$ composing the curve
as follows (Riemann--Hurwitz theorem):
\eqn\rihu{2g-2 = -2n + \sum_{s=1}^L (\nu_s -1)}
The genus $g$ can be explicitly computed once the form of the Weierstrass
polynomial is known exploiting the Baker's method, see
ref. \forsyth . This will be done at the end of this Section.

On a Riemann surface $\Sigma_g$ represented as an $n-$sheeted cover of ${\bf
CP}^1$ there is a ``canonical'' complex structure inherited from
${\bf CP}^1$. A possible atlas on $\Sigma_g$ is the following.
Let us put $R={\rm max}|a_i|$ and $\rho={\rm min}|a_i-a_j|$ for
$i,j=1,\ldots,N_{bp}$. Near a branch point $a_i$ of ramification index
$\nu_i$, or more precisely in the open disk $|z-a_i|<\rho$, we choose
the local coordinate $\xi^{\nu_i}=z-a_i$.
For $|z|>R$, the local coordinate is $z'=1/z$. Let us notice
that on the algebraic curve the set $|z|>R$ corresponds to an union of
$n$ disjoint discs.
On the remaining open sets the
local coordinate is $z$ (the same letter $z$ denotes here coordinates on
$n$ different branches of $\Sigma_g$;
this convention is very useful and, hopefully, does not
generate ambiguities).

In the rest of this Section, we discuss the construction
of the relevant meromorphic tensors and the computation of their
divisors. We are interested in tensors of the kind
$T^{(\alpha)}(z)dz^\lambda$, with $\lambda$ upper or lower indices
depending on the sign of $\lambda=0,\pm 1,\pm 2,\ldots$.
The treatment of
tensors characterized by half-integer values of $\lambda$
is possible
only on hyperelliptic curves and will not be considered here.
The meromorphic functions correspond to the case $\lambda=0$.
The branch index $\alpha$ has been added to recall that a tensor $T$ is
multivalued on ${\bf CP}^1$ due to its
dependence on $y(z)$.
To any meromorphic tensor $Tdz^\lambda$ with zeros at $z_r$ of order $k_r$
and poles at $p_s$ of order $l_s$, one can associate a divisor $[T]$
\fakra\ :
\eqn\div{[T] = \sum_r k_rz_r - \sum_s l_sp_s.}
$k_r$ and $l_s$ are integers while $z_r$ and $p_s$ denote points
on the algebraic curve
The degree of the divisor $[T]$ is defined as follows:
\eqn\divdeg{deg[T]= \sum_r k_r - \sum_sl_s.}
The most general tensor on an algebraic curve is of the form:
\eqn\protodiff{T^{(\alpha)}(z)dz^\lambda= Q(z,y^{(\alpha)}(z)){dz^\lambda\over
[F_y(z,y^{(\alpha)}(z))]^\lambda}}
where $Q(z,y)$ is a rational function of $z$ and $y$.
The reason for which the factor $[F_y(z,y^{(\alpha)}(z))]^{-\lambda}$
has been singled out in \protodiff\
will soon become clear.
From eq. \protodiff\ it is evident that, in order to construct tensors
on an algebraic curve with poles and zeros at given points, it is necessary
to know the divisors of the basic building blocks $dz$, $y$ and
$F(z,y)$.
This can be done quite explicitly for the
general polynomials described by eq. \weipd\ if
$P_n(z)=1$ and if there are no branch points at infinity.
We only need the additional assumption that the polynomials
$P_1(z)$ and $P_0(z)$ appearing in $F(z,y)$ have no roots in common.
In this way, 
eq. \curve\ is approximated  for small values of  $y$
by the
relation
$y\sim -P_0(z)/P_1(z)$.
Therefore, the zeros $q_1,\ldots,q_{n_0}$ of $y(z)$ occur for values of
$z$ corresponding to the roots of
$P_0(z)$.
To study the behavior of $y(z)$ at infinity we try the ansatz
\ansatz\ in eq. \curve. If we retain only the leading order terms
of $y(z)$ and of the polynomials $P_s(z)$ appearing in \weipd,
then eq. \curve\ is approximated by:
\eqn\leaord{c^n
z^{pn}+\ldots+A_{s,n_s}c^{s}z^{ps+n_s}+
\ldots+A_{0,n_0}z^{n_0}=0 }
Since by assumption $y(z)$ is not branched
at infinity, there should be
$n$ different solutions for  $c$ that satisfy \leaord.
Clearly, this can be true only if 
the first and last monomials $c^nz^{pn}$ and $A_{0,n_0}z^{n_0}$
entering in eq. \leaord\
are growing with the same power of $z$ near the point
$z=\infty$, i. e. $z^{pn}\sim z^{n_0}$.
Moreover, all the other leading order
monomials appearing in eq. \leaord\
must not contain higher order powers in $z$,
i. e.:
\eqn\condone{
ps+n_s\le n_0\qquad\qquad\qquad s=1,\ldots,n-1}
Thus, we obtain for $p$ the following result:
\eqn\aku{p={n_0\over n}=1,2,\ldots}
This implies that the integer $n_0$ is a multiple of $n$.

In this way we have derived the divisor of $y$:
\eqn\divy{ [y]=\sum^{n_0}_{r=1}q_r\ -\ \sum^{n-1}_{j=0}{n_0\over n}\infty_j }
The symbols $\infty_j$ denote the points on the curve
corresponding to the point
$z=\infty$ in {\bf CP}$_1$.
In the covering of the algebraic curve described above they belong to the $n$
disjoint discs where the condition $|z|>R$ is satisfied.
As we see from the above equation, the degree of the divisor $[y]$ is
zero as expected for a meromorphic function.
Analogously, it is
possible to compute the divisors of $F_y(z,y)$ and $dz$:
\eqn\divdf{[F_y]=\sum^{n_{bp}}_{r=1}(\nu_{r}-1)a_r\ -\
(n-1)\sum^{n-1}_{j=0}{n_0\over n}\infty_j }
\eqn\divdz{ [dz]=\sum^{n_{bp}}_{r=1}(\nu_r-1) a_r\ -\
2\sum^{n-1}_{j=0}\infty_j }
The details are explained in ref. \naszed.
Exploiting the above divisors, one is able to prove that,
if $\lambda\ge 2$, the following tensor has only a simple pole at the
point $z=w$ on the sheet $\alpha=\beta$:
\eqn\weiker
{
K_{\lambda}^{(\alpha\beta)} (z,w)dz^\lambda = {1\over z-w} 
{F(w,y^{(\alpha)}(z))\over [y^{(\alpha)}(z) - y^{{(\beta)}}(w)]}
{dz^\lambda
\over [F_y(z,y^{(\alpha)}(z))]^{\lambda} }}
where the indices $\alpha$ nd $\beta$  label the branches in $z$ and $w$
respectively.
The tensor $K_{\lambda}^{(\alpha\beta)} (z,w)dz^\lambda $ will
be hereafter called
the Weierstrass kernel.
If $\lambda =1$, it is easy to check that
\eqn\ake{\nu_{uw}^{(\alpha\beta\gamma)}(z)dz = K_1^{(\alpha\beta)}(z,u)dz-
K_1^{(\alpha\gamma)} (z,w)dz}
is a differential of the third kind \foot{Following
the classification of
\ref\griff{P. A. Griffiths,  {\it Introduction to
Algebraic Curves}, Translations of Mathematical
Monographs, Vol. 76, American Mathematical Society, Providence,
Rhode Island, 1989}, here third kind differentials
are defined as meromorphic differentials with at most simple
poles. This definition differs for instance from
that of \fakra.}
with two simple
poles in $z=u$ and $z=w$
on the sheets $\alpha=\beta$ and $\alpha=\gamma$ respectively. 
The residue of $\nu_{uw}^{(\alpha\beta\gamma)}(z)dz$
is $+1$ at $z=u$ and $-1$ at $z=w$.
For our purposes we will also need differentials of the first kind,
or holomorphic one
forms and differentials of the second kind, which consists
of meromorphic one
forms with vanishing residue.
Any meromorphic differential can
be decomposed in terms of the elements of a generalized Laurent
basis given by \naszejdwa:
\eqn\eleone{f_{k,i}(z)dz = {z^{-i-1}y^{n-1-k}(z)\over F_y(z,y(z))}dz
\qquad\qquad\qquad \cases{i=0,\pm 1,\pm2,\ldots\cr
k=0,\ldots,n-1\cr}}
For the functions, instead, it is possible to use the
following basis:
\eqn\eletwo{\phi_{k,i}(z)= z^{-i}\left(y^{k}(z)+y^{k-1}(z)P_{n-1}(z)
+y^{k-2}(z)P_{n-2}(z)+\ldots+P_{n-k}(z)\right)}
where $k$ and $i$ take the same values as in eq. \eleone.
The elements of the basis \eletwo\ look apparently complicated, yet they are
very convenient in order to expand the differentials of the
third kind \ake.
As a matter of fact, it is possible to show that \naszed:
\eqn\dtkexp{
\nu_{uw}^{(\alpha\beta\gamma)}(z)dz=
\sum_{k=0}^{n-1}
{f^{(\alpha)}_{k,-1}(z)\phi^{(\beta)}_{k,0}(u)\over z-u}dz
-\sum_{k=0}^{n-1}
{f^{(\alpha)}_{k,-1}(z)\phi^{(\gamma)}_{k,0}(w)\over z-w}dz
}

To construct a basis of holomorphic differentials it is sufficient to find
all possible values of $s$ and $k$ in eq. \eleone\ for which
$f_{k,s}(z)dz$ has no poles.
It is easy to check with the help of eqs. \divy--\divdz\
that such a basis is given by:
\eqn\holodiff{\Omega_{(j,s_j)}(z)dz={z^{s_j}y^j(z)\over F_y(z,y(z))}dz}
where, for $p=1,2,\ldots$
\eqn\jsjdefs{j=0,\ldots,n-2-\delta_{p,1}\qquad\qquad
s_j=0,\ldots,(n-1-j)p-2\qquad\qquad\delta_{p,1}=\cases{1\enskip{\rm if}
\enskip p=1\cr
0\enskip{\rm if}
\enskip p>1\cr}}
The number of the
above holomorphic differentials coincides with the genus of the curve
\weipd.
A straightforward computation gives:
\eqn\genusholo{g={pn(n-1)-2(n-1)\over 2}}
The above calculation of $g$ can be verified using the already mentioned
method of Baker.
To this purpose, let us consider a two dimensional integer
lattice on the $x-y$ plane.
One draws a triangle $OAC$ with vertices at the points
$O=(0,0)$, $A=(a,0)$ and $C=(0,b)$, where $a=n$ and $b=pn$.
According to Baker's method, the genus of the curve \weipd\
with the additional constraints \condone--\aku\ and $P_n(z)=1$ coincides
with the number $\tilde g$ of lattice points contained inside the area of this
triangle, the boundary excluded.
Let us evaluate $\tilde g$. First of all, one counts the number of
lattice points $T$ inside
the rectangle $OABC$, where $B=(a,b)$, excluding also the boundary.
Then one subtracts the number of links $D$
lying on the diagonal $AC$.
One finds that $T=(pn-1)(n-1)$ and $D=n-1$.
Clearly
\eqn\genustilde{\tilde g={T-D\over 2}}
and it is easy to see that this gives for the genus of the
curve \weipd\ exactly the result of eq. \genusholo, i. e.
$\tilde g=g$.

Finally, a possible metric on the curve \weipd\
is given by
\eqn\metric{
ds^2 =\tilde
\rho_{z\bar z}dzd\bar z=(1+z\bar z)^\mu{dzd\bar z\over |F_y(z,y)|^2}\qquad
\qquad\qquad \mu = p(n-1) - 2}

\newsec{SCALAR GREEN FUNCTIONS ON ALGEBRAIC CURVES}

In this Section we consider the free bosonic scalar field theory
described by the action:
\eqn\scalaction{S=\int_{\Sigma_g}d^2 z\partial\varphi\bar \partial\varphi}
defined on a general algebraic curves
$\Sigma_g$ discussed above. A conformal metric on $\Sigma_g$ is 
understood in \scalaction\ . Moreover the $d^2z$ is a shorthand notation for 
${1\over 2i} dzd\bar z$.

A scalar field in the  two dimensional field theory \scalaction\ 
satisfies the Poisson equation
can can be interpreted as an electrostatic
potential of a Coulomb system of charges. 
$\partial_z\varphi dz$
is a meromorhic 
differential with the sum of residua at poles equal zero. Consequently
the sum of charges on the Riemann surface must be zero as well.
Thus we consider a system of charges $q_i$ set
in the positions
$z_i$, $i=1,\ldots,M$ interacting with another system of charges $q'_j$
located at the points
$w_j$, $j=1,\ldots,N$ and satisfying the relations
$\sum\limits_i q_i=
\sum\limits_j q'_j=0$.
The corresponding correlation function:
\eqn\genbosampl{G(z_1\ldots z_M;w_i\ldots w_N) =
\sum_{i=1}^M\sum_{j=1}^N q_i\langle\varphi(z_i,\bar z_i)\varphi(w_j,\bar w_j)
\rangle q'_j}
can be computed once the following Green function is known:
\eqn\basicgf{G_z(z;u,w)=\langle\partial_z\varphi(z,\bar z)
\left[\varphi(u,\bar u)-\varphi(w,\bar w)\right]\rangle}
As a matter of fact, one can prove the following relation:
$$
G(z_1\ldots z_M;w_i\ldots w_N) =
\sum_{i=1}^M\sum_{j=1}^N q_iq'_j{\rm Re}\left[\int_{z_0}^{z_i}
dz\langle\partial_z\varphi(z,\bar z)
\left[\varphi(w_j,\bar w_j)-\varphi(u,\bar u)\right]\rangle \right]=$$
\eqn\gensimplif{
{1\over 2}\sum_{i=1}^M\sum_{j=1}^N q_iq'_j \left[\int_{z_0}^{z_i}
dz\langle\partial_z\varphi(z,\bar z)
\left[\varphi(w_j,\bar w_j)-\varphi(u,\bar u)\right]\rangle + cc \right]}
Moreover, once $G_z(z;u,w)$ is known, one obtains also the
correlator
\eqn\gzu{G_{zu}(z;u) =\langle\partial_z\varphi(z,\bar z)\partial_u
\varphi(u,\bar u)\rangle}
as can be seen
by differentiating
both sides of equation \basicgf\ with respect to $u$.
For these reasons, we will limit ourselves to the computation of
$G_z(z;u,w)$.

To simplify the notations, we introduce the composite indices
$I,J,K,\ldots$. with $I=(i,s_i)$, $J=(j,s_j)$ etc.
For instance, in this notation $\Omega_I(z)dz=\Omega_{(i,s_i)}(z)dz$.
It is now possible to construct
$G_z(z;u,w)$ starting from any third kind differential with two poles
at $z=u,w$ and adding to it a suitable
linear combination of holomorphic differentials.
Using for example the third kind differential defined in eq.
\ake, we have in general:
\eqn\gzcoeffdef{G_z(z;u,w)dz=
\nu_{uw}(z)dz+\sum_I B_I(u,w)\Omega_I(z)dz}

Following the strategy of \ref\ijmpen{F. Ferrari,
{\it Int. Jour. Mod. Phys.} {\bf A5} (1990), 2799.}, we try to 
determine the coefficients $B_I(u,w)$ exploiting the fact that
the fields $\varphi(z,\bar z)$ are single-valued functions
on $\Sigma_g$, so that they must obey the relations:
\eqn\requirone{\oint_{\gamma_s}d\varphi=\oint_{\gamma_s}\partial_z\varphi dz
+\oint_{\bar\gamma_s}\bar\partial_{\bar z}\varphi d\bar z=0}
where the $\gamma_s$, $s=1,\ldots,2g$, form a basis of
homology cycles on $\Sigma_g$.
For consistency, the right hand
side of eq. \gensimplif\ has to be single-valued along the homology cycles.
This condition is satisfied if and only if
the correlator $G_z(z;u,w)$ is a differential of
the third kind normalized in such a way
that all its
periods are purely imaginary. Let us denote this normalized differential
with the symbol
$\omega_{uw}(z)dz$. Thus we require that:
\eqn\normtkdiff{G_z(z;u,w)dz=\omega_{uw}(z)dz}
The above equation determines $G_z(z;u,w)dz$ uniquely.
Comparing in fact eq. \gzcoeffdef\ with the above equation
and imposing the condition that
$\nu_{uw}(z)dz+\sum_I B_I(u,w)\Omega_I(z)dz$ has imaginary periods
around all homology cycles,
one obtains the following linear system of $2g$ equations
in the $2g$ real unknowns ${\rm Re}[B_I(u,w)]$ and
${\rm Im}[B_I(u,w)]$:
\eqn\linsys{{\rm Re}\left[
\oint_{\gamma_s}\nu_{uw}(z)dz+\sum_I
B_I(u,w)\oint_{\gamma_s}\Omega_I(z)dz
\right]=0\qquad\qquad s=1,\ldots,2g}
Unfortunately, the construction of a basis of homology cycles
on algebraic curves of the kind discussed here is a complicated mathematical
problem. As a consequence,
the periods $\oint_{\gamma_s}\nu_{uw}(z)dz$
and $\oint_{\gamma_s}\Omega_I(z)dz$ cannot be evaluated in a closed form.
An exception is provided by the $Z_n$ symmetric
curves, where these periods are
expressed as definite integrals having the branch points as extrema.
In the present case, however, even the positions of the branch points
is unknown. The reason is that they are the roots of the polynomial
$r(z)=0$, where $r(z)$ is the resultant of the system
\bpdef\ as discussed in the previous Section.
Clearly, those roots cannot be computed analytically apart from
simple cases.

To avoid these difficulties, we evaluate the coefficients
$B_I(u,w)$ using the Riemann's bilinear equations \fay\ :
\eqn\rbileqs{0=\int_{\Sigma_g}\omega_{uw}\wedge\bar\Omega_J=
\int_{\Sigma_g}d^2z\omega_{uw}(z)\bar\Omega_I(\bar z)}
where the $\bar\Omega_J(\bar z)$'s denote the antiholomorphic differentials
on $\Sigma_g$.
In Appendix A we will show
that the above relation holds if and only if
$\omega_{uw}(z)dz$ is a normalized differential of the third kind
with imaginary periods along the nontrivial homology
cycles of $\Sigma_g$.

The integral in \rbileqs\ is 
to be understood as a sum over all possible
values of $z$ and $y$ for which the relation
$F(z,y)=0$ is satisfied.
Supposing that the curve $\Sigma_g$ is nondegenerate,
eq. \rbileqs\ can be put in a form which is more convenient for explicit
computations:
\eqn\intexpl{I=\int_{{\rm\bf CP}^1}d^2z
\sum_{\alpha=0}^{n-1}f_{z\bar z}
(z,\bar z;y^{(\alpha)}(z),\overline{y^{(\alpha)}(z)})}
A detailed proof of eq. \intexpl\ is provided in Appendix B
(see also \grha).
Once the polynomial $F(z,y)$ is given, it is possible to compute
the integral \intexpl\ numerically.
As a matter of fact, to evaluate the integrand
\eqn\integrand{I_{z\bar z}(z,\bar z)=
\sum_{\alpha=0}^{n-1}f_{z\bar z}
(z,\bar z;y^{(\alpha)}(z),\overline{y^{(\alpha)}(z)})}
appearing in \intexpl\ at any given point $z\in{\rm\bf CP}^1$,
we just need to invert the equation $F(z,y)=0$.
The latter is a polynomial equations
of degree $n$ from which one derives the $n$ roots
$y^{(0)}(z),\ldots,y^{(n-1)}(z)$. The values of
$\overline{y^{(0)}(z)},\ldots,\overline{y^{(n-1)}(z)}$
are obtained by complex conjugation.
The advantage of this strategy is that all branches
of $y(z)$ enter symmetrically in the sum of eq. \integrand.
Thus it is not necessary to
know how these branches are exchanged at the branch points
or any other information which requires the analytic continuation of
$y(z)$.

Using the identity \normtkdiff, we substitute in eq.
\rbileqs\ the expression
of $G_z(z;u,w)dz$ in terms of the not normalized
differential $\nu_{uw}(z)dz$ given by eq.
\gzcoeffdef.  The solution of the obtained in this way linear 
equations for coefficients $B_I(u,w)$:

\eqn\rbileqs{
0=\int_{\Sigma_g}
\nu_{uw}\wedge\bar\Omega_J+\sum_I B_I(u,w)
\int_{\Sigma_g}\Omega_I(z)\wedge
\bar\Omega_J}

is in the matrix form given by

\eqn\sobczyk{B_I(u,w)=-\sum_J\left[\int_{\Sigma_g}
\nu_{uw}\wedge\bar\Omega_J\right]\left[\int_{\Sigma_g}\Omega_J(z)\wedge
\bar\Omega_I\right]^{-1}}

In compact form the final result can be written up as ($\alpha ,\beta ,
\gamma$ label sheets of the Riemann surface):
\eqn\gzfinal{
G_z^{(\alpha\beta\gamma)}(z;u,w)=
{1\over \det A_{IJ}}
\det
\left(\left.
{
\raise4pt\hbox{$\nu_{uw}^{(\alpha\beta\gamma)}(z)$}
\over
\lower8pt\hbox{$\Phi_J^{(\alpha)}(u) - \Phi_J^{(\beta)}(w)$}}
\right|
\left.
{
\raise4pt\hbox{$\Omega_I^{(\alpha)}(z)$}
\over
\lower8pt\hbox{$A_{IJ}$}
}\right.\right)}
where
\eqn\holoints{
A_{IJ}\equiv A_{(i,s_i)(j,s_j)}=
\int_{{\rm\bf CP}^1}d^2z\sum_{\alpha=0}^{n-1}{
\left[y^{(\alpha)}(z)\right]^i\overline{\left[y^{(\alpha)}(z)\right]}^j
\over\left|F_y(z,y^{(\alpha)}(z))\right|^2}z^{s_i}\bar z^{s_j}}
and
\eqn\tkdints{
\Phi_J^{(\beta)}(u)\equiv\Phi_{(j,s_j)}^{(\beta)}(u)=
\sum_{k=0}^{n-1}\phi_{k,0}^{(\beta)}(u)
\sum_{\alpha=0}^{n-1}\int_{{\rm\bf CP}^1}d^2z
{\left[y^{(\alpha)}(z)\right]^{n-1-k}
\overline{\left[y^{(\alpha)}(z)\right]}^j\bar z^{s_j}
\over \left|F_y(z,y^{(\alpha)}(z))\right|^2(z-u)}
}
In eq. \holoints\ the range of the integers $i,j,s_i$ and $s_j$
is given by eq. \jsjdefs.
Moreover, we notice that the integrals in eqs. \holoints\ and \tkdints\
are convergent, since the integrands have at most harmless  poles
of the first order.
These singularities occur only in the integrals \tkdints\ at the
points $z=u,w$ and at infinity. The singularity at infinity
is present only when $k=0$ and $p>1$. Clearly, the discussion
following eq. \intexpl\ applies also to the particular
case of the integrals in
\holoints\ and \tkdints, which can thus be computed numerically
without problems of analytical continuation.

Eqs. \gzfinal-\tkdints\ provide an explicit representation
of the correlator $G_z(z;u,w)$, where neither the knowledge
of a basis of homology cycles nor the analytic continuation
of multivalued functions are necessary.
This is a great advantage, since the latter are formidable problems
in the case of general curves like those discussed in this paper.
Of course, the integrals
\holoints-\tkdints\ are complicated and cannot be computed analytically,
but only numerically, exploiting for instance
the recipe explained after eq. \intexpl.

To conclude this Section, we study the behavior of
$G_z(z;u,w)dz$ for large values of $u$, showing that there are no spurious
divergences in this case.
From eqs. \dtkexp\ and \tkdints\ one realizes that
$G_z(z;u,w)dz$ depends on $u$ through the function
\eqn\fofu{f(u)={\phi_{k,0}(u)\over z-u}}
When $u$ becomes large, one can write:
\eqn\fofusplit{f(u)=f^{\rm div}(u)+f^{\rm fin}(u)}
where $f^{\rm div}(u)$ diverges in $u=\infty$ while
$f^{\rm fin}(u)$ remains finite.
It is easy to see that:
\eqn\fofudiv{f^{\rm div}(u)=-{\phi_{k,0}(u)\over u}\sum_{i=0}^{pk-2}\left(
{z\over u}\right)^i}
with
\eqn\kgzpgo{k>0\qquad {\rm if}\qquad p>1}
and
\eqn\kgopeo{k>1\qquad {\rm if}\qquad p=1}
Using the above formula, we extract the diverging contributions
$G_z^{\rm div}(z;u,w)dz$ which are present in the correlator $G_z(z;u,w)dz$.
To this purpose, it is convenient to define the composite index:
\eqn\cmpsidx{\overline{I}(k,i)=(n-1-k,i)}
Hereafter, the dependence of $\overline{I}$ on $k$ and $i$
will be only understood for simplicity.
After some calculations one finds:
\eqn\calcb{G_z^{\rm div}(z;u,w)=
{1\over \det A_{IJ}}
\sum_{k=1+\delta_{p,1}}^{n-1}\sum_{i=0}^{pk-2}
{\phi_{k,0}(u)\over u}
\det
\left(\left.
{
\raise4pt\hbox{$\Omega_{\overline{I}}(z)$}
\over
\lower8pt\hbox{$A_{\overline{I}J}$}}
\right|
\left.
{
\raise4pt\hbox{$\Omega_K(z)$}
\over
\lower8pt\hbox{$A_{JK}$}
}\right.\right)}
Clearly
\eqn\vanishing{\det
\left(\left.
{
\raise4pt\hbox{$\Omega_{\overline{I}}(z)$}
\over
\lower8pt\hbox{$A_{\overline{I}J}$}}
\right|
\left.
{
\raise4pt\hbox{$\Omega_K(z)$}
\over
\lower8pt\hbox{$A_{JK}$}
}\right.\right)=0}
showing that $G_z(z;u,w)dz$ has no singularities for large values of
$u$ as desired.
The absence of divergences when $w$ approaches infinity can be proved
in the same way.

\newsec{EXAMPLES}

A first non-trivial example of curves
which can be explicitly worked out is provided
by the algebraic curve:
\eqn\gfour{y^3+3py-2q=0}
$q(z)$ is a polynomial of degree $n_0=6$, while $p(z)$ has degree
$n_1\le{2n_0\over 3}=4$. Thus, exploiting the formulas of Section 2, one realizes
that the curve \gfour\ corresponds to the particular case $p=2$ and $g=4$.
For future convenience let us put:
\eqn\xipmdef{\xi_\pm(z)=
\root3\of{q\pm\sqrt{q^2+p^3}}}
Solving equation \gfour\ with respect to $y$ one finds:
\eqn\ygtbranches{y^{\alpha}=\epsilon^\alpha \xi_+ +
\epsilon^{2\alpha}\xi_-\qquad\qquad\alpha=0,1,2\qquad\qquad
\epsilon=e^{2\pi i\over 3}
}
The four holomorphic differentials of the above curve are:
\eqn\gfholodiff{
\Omega_{(0,s_0)}(z)dz={z^{s_0}dz\over F_y}\qquad\qquad
\Omega_{(1,1)}(z)dz={ydz\over F_y} }
where $s_0=0,1,2$ and $F_y=3(y^2+p)$.
The differential of the third kind of eq. \ake\ becomes in this particular
case:
\eqn\gftkd{\nu_{uw}(z)dz={y^2(z)+y(z)y(u)+y^2(u)+3p(u)\over F_y(z,y(z))(z-u)}
-(u\leftrightarrow w)}
At this point, we are ready to compute the matrix elements $A_{IJ}$ and
$\Phi_I$.
They will be expressed as integrals of single-valued forms on {\bf CP}$^1$.
To simplify the notations, let us first define the following functions:
\eqn\calffun{{\cal F}(z)=\prod_{\alpha=0}^2F_y(z,y^{(\alpha)}(z))}
\eqn\safun{a(z,\bar z)=\xi_+^2(z)\overline{\xi^2_+(z)}+\xi_-^2(z)
\overline{\xi_-^2(z)}+p(z)\overline{p(z)}}
\eqn\sbfun{b(z,\bar z)=\xi_+^2(z)\overline{\xi_-^2(z)}-p(z)
\overline{\xi^2_+(z)}-
\overline{p(z)}\xi_-^2(z)}
\eqn\scfun{c(z,\bar z)=\overline{\xi^2_+(z)}\xi_-^2(z)-p(z)
\overline{\xi_-^2(z)}
-\overline{p(z)}\xi^2_+(z)}
\eqn\afun{A(z,\bar z)=3(a^2-bc)}
\eqn\bfun{B(z,\bar z)=3\left[\xi_+\left(b^2-ac\right)+\xi_-\left(c^2-ab\right)
\right]}
\eqn\cfun{C(z,\bar z)=\left[\overline{\xi_+}\left(c^2-ab\right)+
\overline{\xi_-}\left(b^2-ac\right)\right]}
\eqn\dfun{D(z,\bar z)=3\left[\left(\xi_+\overline{\xi_+}+\xi_-\overline{\xi_-}
\right)(a^2-bc)+\xi_-\overline{\xi_+}(b^2-ac)+\xi_+\overline{\xi_-}(c^2-ab)
\right]}
\eqn\efun{E(z,\bar z)=3\left[2\xi_+\xi_-(a^2-bc)+\xi^2_-(b^2-ac) +
\xi_+^2(c^2-ab)\right]}
$$F(z,\bar z)=$$
\eqn\ffun{3\left[\left(\xi_-^2\overline{\xi_+}+\xi_+^2
\overline{\xi_-}\right)(a^2-bc)+ \left(\xi_+^2\overline{\xi_+}-2p
\overline{\xi_-}\right)(b^2-ac)+\left(\xi_-^2\overline{\xi_-}-2p
\overline{\xi_+}\right)(c^2-ab)\right]}
The above functions are not meromorphic, since they
depend both on $z$ and $\bar z$, but are single-valued on {\bf CP}$_1$.
To prove this fact, we notice that under
the most general monodromy transformation for going from one branch of the curve to another  $\xi_+$
behaves as
\eqn\transfxp{\xi_+(z)\rightarrow
\epsilon^\alpha\xi_\pm(z)\qquad\qquad\alpha=0,1,2\enskip{\rm mod}\enskip 3}
as can be easily seen from the explicit expression of $\xi_+$ given by
eq. \xipmdef. 
The above equation fixes also the monodromy properties of $\xi_-(z)$.
In fact, since $\xi_+(z)$ and $\xi_-(z)$ must satisfy the
relation $\xi_+(z)\xi_-(z)=-p(z)$, where $p(z)$ is a
singlevalued function of $z$, it is clear from eq. \transfxp\ that
$\xi_-(z)$ should transform as follows:
\eqn\transfxm{\xi_-(z)\rightarrow\epsilon^{2\alpha}\xi_\mp(z)}
Moreover, taking the complex conjugate of both members of eqs.
\transfxp\ and \transfxm, one finds:
\eqn\transfxbp{\overline{\xi_+(z)}\rightarrow\epsilon^{2\alpha}
\overline{\xi_\pm(z)}}
\eqn\transfxbm{\overline{\xi_-(z)}\rightarrow\epsilon^\alpha
\overline{\xi_\mp(z)}}
It is now easy to check that the functions defined in
eqs. \afun--\ffun\ are invariant
under the transformations \transfxp--\transfxbm, so that they are
single-valued on {\bf CP}$^1$.
In terms of these functions,
the matrix elements $A_{IJ}$ and
$\Phi_I$ appearing in the normalized differential of the third kind
\gzfinal\ read  as follows:
\eqn\azszazaz{A_{(0,s_0)(0,s_0')}=\int_{{\rm\bf CP}^1}d^2z{z^{s_0}
\bar z^{s_0'}\over\left|{\cal F}(z)\right|^2}A(z,\bar z)}
\eqn\aozzsz{A_{(1,0)(0,s_0)}=\int_{{\rm\bf CP}^1}d^2z{
\bar z^{s_0}\over\left|{\cal F}(z)\right|^2}B(z,\bar z)}
\eqn\zszoz{A_{(0,s_0)(1,0)}=\int_{{\rm\bf CP}^1}d^2z{
z^{s_0}\over\left|{\cal F}(z)\right|^2}C(z,\bar z)}
\eqn\ozoz{A_{(1,0)(1,0)}=\int_{{\rm\bf CP}^1}d^2z{
D(z,\bar z)\over\left|{\cal F}(z)\right|^2}}
$$\Phi_{(0,s_0)}(u)=\int_{{\rm\bf CP}^1}d^2z{
\bar z^{s_0}E(z,\bar z)\over\left|{\cal F}(z)\right|^2(z-u)}+$$
\eqn\pzsz{y(u)\int_{{\rm\bf CP}^1}d^2z{
\bar z^{s_0}B(z,\bar z)\over\left|{\cal F}(z)\right|^2(z-u)}+
(y^2(u)+3p(u))\int_{{\rm\bf CP}^1}d^2z{
\bar z^{s_0}A(z,\bar z)\over\left|{\cal F}(z)\right|^2(z-u)}}
$$\Phi_{(1,0)}(u)=\int_{{\rm\bf CP}^1}d^2z{
F(z,\bar z)\over\left|{\cal F}(z)\right|^2(z-u)}+$$
\eqn\poz{y(u)\int_{{\rm\bf CP}^1}d^2z{
D(z,\bar z)\over\left|{\cal F}(z)\right|^2(z-u)}+
(y^2(u)+3p(u))\int_{{\rm\bf CP}^1}d^2z{
C(z,\bar z)\over\left|{\cal F}(z)\right|^2(z-u)}}
Substituting the matrix elements \azszazaz--\poz\ and the third kind
differential \gftkd\ in eq. \gzfinal, we obtain the explicit
expression of the
correlator \basicgf.
As we see, all the integrands appearing
in the right hand sides of eqs. \azszazaz--\poz\ are one-valued on the
complex sphere as desired,
since they depend on the one-valued functions \calffun--\ffun.
In this way the integrals in \azszazaz--\poz\ can be numerically evaluated
without the problem of performing the analytic continuation of $y(z)$.
Also the problem of constructing a basis of homology cycles has been avoided.

Let us now briefly discuss
the simple case of the $Z_n$ algebraic curves
of the kind:
\eqn\znsymm{y^n=p_0(z)}
where $p_0(z)$ is a polynomial of degree $n_0=n,2n,3n,\ldots$.
The genus of these curves is provided by the general formula \genusholo.
For $p>1$, i.e. $n_0>n$, the zero modes are:
\eqn\znzeromod{\Omega_{j,s_j}(z)dz={z^{s_j}y^j(z)\over y^{n-1}}dz}
The matrix elements $A_{IJ}$ are given by:
\eqn\znzmz{A_{(i,s_i)(j,s_j)}=0\qquad\qquad i\ne j}
\eqn\znzmnz{A_{(j,s_j)(j,s_j)}=n\int_{{\rm\bf CP}^1}d^2zz^{s_j}\bar z^{s_j}
|y\bar y|^{1-n+j}}
Clearly, the integrand in the right hand side of the above equation
is single-valued.
In the same way one can discuss the elements $\Phi_I(u)$.
This is a straightforward exercise which will not be performed here.
Instead, we compute on a $Z_n$ symmetric curve the propagator
\eqn\znfprop{G(z;a)=\langle\varphi(z,\bar z)\varphi(a,\bar a)\rangle}
where $a$ is one of the $pn$ branch points of the curve \znsymm.
The starting point is the Weierstrass kernel at a branch point, which
in the case of a $Z_n$ symmetric 
curve takes a particularly simple form  (put $y(a)=0$ in \weiker\ ):
\eqn\gznbp{K_1(z,a)dz={1\over n}{dz\over z-a}}
We define the following function on the complex sphere:
\eqn\kzai{{\cal G}(z,w)={1\over 2n}{\rm log}\left({|z-w|^2\over (1+z\bar z)}
\right)}
which has only logarithmic singularities for $z\rightarrow \infty$.
${\cal G}(z,w)$
satisfies the identity:
\eqn\kzweq{\partial_z\partial_{\bar z}{\cal G}(z,w)={\pi\over n}
\delta^{(2)}_{z\bar z}(z-w)
-{1\over 2n}{1\over (1+z\bar z)^2}}
where $\delta^{(2)}_{z\bar z}(z-w)$ is defined by
\eqn\dirac{\int_{{\bf CP}^1}d^2z
f(z,\bar z)\delta^{(2)}_{z\bar z}(z-w)=f(w,\bar w).}
\kzweq\ is the relation satisfied by a Green function on the complex
sphere with the metric $ds^2={dzd\bar z\over(1+z\bar z)^2}$.
Let us now consider a two-form $f(z,\bar z)=f_{z\bar z}(z,\bar z){d\bar 
z\wedge z\over 2i}$ on $\Sigma_g$ with 
sufficiently smooth behavior.
$f_{z\bar z}$ is in general
a multivalued function on {\bf CP}$_1$. Exploiting
eq. \intexpl, we have:
$$
\int_{\Sigma_g}d^2wf_{w\bar w}(w,\bar w;y,\bar y)
\tilde\delta^{(2)}_{z\bar z}(z-w)\equiv \sum_{\alpha=0}^{n-1}\int_{{\rm\bf CP}_1}d^2w
f_{w\bar w}(w,\bar w;y^{(\alpha)}(w),\overline{y^{(\alpha)}(w)})
\delta^{(2)}_{z\bar z}(z-w)=$$
\eqn\ddd{
=\sum_{\alpha=0}^{n-1}
f_{z\bar z}^{(\alpha)}(z,\bar z).}
On $Z_n$ symmetric curves near branch points all the branches of $y(z)$
coincide.
For this reason, if $z\rightarrow a$ eq. \ddd\ becomes:
\eqn\ddfnabp{\int_{\Sigma_g}d^2wf_{w\bar w}(w,\bar w;y,\bar y)
\tilde\delta^{(2)}_{z\bar z}(a-w)=
n\left.f_{z\bar z}(z,\bar z)\right|_{z=a\atop
\bar z = \bar a}}
Thus, the proper definition of a
$\delta-$function on $\Sigma_g$ at a branch point is ${1\over n}
\tilde\delta_{z\bar z}^{(2)}(z-a)$.

We
try now an ansatz
\eqn\ansatzgza{G(z;a)={\cal G}(z,a)-{1\over A}\int_{\Sigma_g}d^2w
\tilde\rho_{w\bar w}{\cal G}(z,w)}
for the propagator \znfprop. In the obvious way 
${\cal G}$ is treated in \ansatzgza\ as an object on $\Sigma_g$
The metric $\tilde\rho_{w\bar w}$ is given by eq. \metric, while
\eqn\znarea{A=\int_{\Sigma_g}d^2w \tilde\rho_{w\bar w}}
is the area of $\Sigma_g$.
We notice that, by construction, the metric
$\tilde\rho_{w\bar w}$
is $Z_n$ symmetric, so that all its branches coincide. As a consequence:
\eqn\mtz{\tilde\rho_{w\bar w}(w,\bar w;y^{(\alpha)}(w)\overline{
y^{(\alpha)}(w)})=\rho_{w\bar w}(w,\bar w)}
where $\rho_{w\bar w}(w,\bar w)$ is a tensor on
{\bf CP}$_1$.
Thus, exploiting again eq. \intexpl, we can put:
\eqn\areasim{A=n\int_{{\rm\bf CP}_1}d^2w \rho_{w\bar w}(w,\bar w)}
Applying now the Laplacian on $\Sigma_g$ 
to both members of eq. \ansatzgza\ and using
eq. \kzweq, we find that $G(z,a)$ satisfies the following identity:
\eqn\iddd{
\partial_z\partial_{\bar z}G(z,a)=\pi\tilde\delta^{(2)}_{z\bar z}(z-a)
-{1\over A}\tilde\rho_{z\bar z}(z,\bar z)}
As explained above, the $\delta-$function on the algebraic curve
near a branch point $a$ is exactly ${1\over n}\delta^{(2)}_{z\bar z}(z-a)$.
Moreover, all the branches of the metric $\tilde\rho_{z\bar z}$
coincide with $\rho_{z\bar z}(z,\bar z)$.
Thus we conclude that eq. \ansatzgza\ is the desired propagator
of the scalar fields when one of the fields is located at a branch point.

\newsec{CONCLUSIONS}

In this paper the
correlator $G_z(z;u,w)dz$ defined in eq. \basicgf\ 
has been computed on the general algebraic curves \weipd.
This Green function plays a fundamental role
in the theory of massless scalar fields, since all other
correlation functions can be written in terms of its derivatives
or integrals as shown by eqs. \gensimplif\ and \gzu.
The
expression of $G_z(z;u,w)dz$ given in eq. \gzfinal\  depends
explicitly on the constant parameters $A_{s,m}$
of the polynomial \weipd\ as desired.
In fact, it contains the Weierstrass
kernel and the holomorphic differentials which have been
derived in eqs. \weiker\ and \holodiff\ respectively.
The calculation of the coefficients $A_{IJ}$ and $\phi^{\alpha}_J(w)$,
in principle more tricky, has been reduced to the evaluation of
the integrals \holoints\ and \tkdints\ on the complex sphere.
The latter may very complicated, especially if the polynomial \weipd\ has
high degree $n$, but they do not hide outstanding technical difficulties
and can be performed
at least numerically
(see eq. \intexpl\
and the following discussion).
Actually, the integrands  appearing in eqs.
\holoints\ and \tkdints\ should also be single-valued, because 
they consists in the sum over all branches of multivalued complex forms.
This fact has been explicitly shown in the particular example of an 
algebraic curve of genus four worked out in
Section 4. However, this example shows also that the
single-valuedness is realized in a very non-trivial
way.
Let us notice that the absence of technical problems
in the computation of $G_z(z;u,w)dz$ was
not granted a priori.
For instance, choosing the set of equations
\linsys\ in order to determine this Green function,
we would have not been able to
compute the necessary line integrals along the homology cycles even
numerically, since it is not
known how to construct an homology basis on general algebraic curves.

Concluding,
the formula \gzfinal, which gives the explicit form
of
the canonical third kind differential
with purely imaginary periods,
is new in the theory of algebraic curves and has many potential
applications beyond the theory of free scalar fields.
Let us remember in fact that this canonical differential
is well known in the mathematical literature and has
already been widely used in physical applications of Riemann surfaces.
Also the propagator at the branch points of eq.
\ansatzgza\ may have some interest in string theories. 
Finally, we have developed in this paper several techniques
which
might be useful for whoever is wishing to work
on general algebraic curves.
\vskip 2cm\noindent
ACKNOWLEDGMENTS
One of the authors would 
like to thank Abdus Salam International Centre for Theoretical Physics
for a hospitality.

\appendix{A}{}

In this Appendix we show that the condition \rbileqs\ is valid if and
only if
$\omega_{uw}(z)dz$ is a differential of the third kind normalized
in such a way that all its periods around the homology cycles are imaginary.
In the proof, it will be convenient to use the representation of
the Riemann surface $\Sigma_g$ as a polygon
$M=\prod_{i=1}^g a_jb_ja^{-1}_jb^{-1}_j$, where
$\{a_i,b_i|i=1,\ldots,g\}$ is a canonical basis of homology cycles \fakra .
Accordingly, the notation will be changed with respect to the rest of the
paper. Let $x,y$ be real coordinates on $M$
and $z=x+iy$, $\bar z = x-iy$.
The basis of canonical holomorphic differentials on $\Sigma_g$ will be denoted
as follows:
\eqn\hdnewdef{\omega_i=\omega_i(z)dz\qquad\qquad\qquad i=1,\ldots,g}
The $\omega_i$ are normalized in such a way that:
\eqn\hdnnewdef{
\oint_{a_i}\omega_j=\delta_{ij}
\qquad\qquad\qquad
\oint_{b_i}\omega_j=T_{ij}}
for $i,j=1,\ldots,g$.
The $T_{ij}=T_{ji}$ are elements of the period matrix $T$.
This is a symmetric $g\times g$ matrix with positive definite
imaginary part.
Since $M$ is simply connected, there exist holomorphic
functions $f_j(z)$ such that
\eqn\holofuncts{df_j(z)=\omega_j(z)dz}
In the above equation $d$ denotes the external derivative
operator such that $d^2=0$.
Analogously, one can define the antiholomorphic differentials
$\bar\omega_j=\bar\omega_j(\bar z)d\bar z$
and functions $\bar f_j(\bar z)$.
Finally, third kind differentials will be denoted with the symbol
$\omega_{PQ}=\omega_{PQ}(z)dz$, with $P,Q\in M$.
We are now ready to prove the following

{\bf Proposition:} The conditions
\eqn\newbigleq{\int_M\omega_{PQ}\wedge\bar\omega_j=0\qquad\qquad\qquad
j=1,\ldots,g}
are verified iff $\omega_{PQ}$ is normalized in such a way that all
its periods are
imaginary, i.e.
\eqn\oconjugprops{\overline{\oint_{a_i}\omega_{PQ}}=
-\oint_{a_i}\omega_{PQ}\qquad\qquad\qquad
\overline{\oint_{b_i}\omega_{PQ}}=
-\oint_{b_i}\omega_{PQ}}

{\bf Proof:} We assume first that eq. \newbigleq\ is true.
Introducing the antiholomorphic functions $\bar f_j(\bar z)$
and exploiting the Stoke's theorem after an integration by parts,
we obtain:
\eqn\aaa{0=
-\int_{\partial M}\bar f_j\omega_{PQ}+
\int_M\bar f_jd\omega_{PQ}}
where $\partial M$ is the boundary of the polygon $M$.
Using the Riemann's bilinear formula \fay:
\eqn\bbb{\int_{\partial M}\bar f_j\omega_{PQ}
=\sum_{l=1}^g\left[\overline{\oint_{a_l}\omega_j}\oint_{b_l}\omega_{PQ}
-\overline{\oint_{b_l}\omega_j}\oint_{a_l}\omega_{PQ}\right]}
and remembering eq. \hdnnewdef, the first term in the right side
of eq. \aaa\ becomes:
\eqn\ccc{\int_{\partial M} \bar f_j \omega_{PQ}
=\oint_{b_j}\omega_{PQ} -\sum_{l=1}^g\overline{T}_{lj}\oint_{a_l}\omega_{PQ}}
The second term gives instead:
\eqn\ddd{\int_M\bar f_j d\omega_{PQ}=\int_M\bar f_j(\bar\partial_{\bar z}
\omega_{PQ})d\bar z\wedge dz}
where $\bar\partial_{\bar z}$ denotes partial derivative with respect to
$\bar z$.
At this point, we notice that:
\eqn\diracapp{\bar\partial_{\bar z}\omega_{PQ}(z)=
\pi\left[\delta^{(2)}(z,P)-\delta^{(2)}(z,Q)\right]}
and
$d\bar z\wedge dz=2idx\wedge dy= 2i d^2x$,
where $\delta^{(2)}(z,w)$ is the Dirac $\delta-$ function
on $M$ defined in such a way that
\eqn\eee{\int_Mf(z,\bar z)\delta^{(2)}(z,w)d^2x=f(w,\bar w)}
Thus:
\eqn\fff{\int_M\bar f_jd\omega_{PQ}= 2\pi i\left(\bar f_j(\bar P)-
\bar f_j(\bar Q)\right)=2\pi i \overline{\int_Q^P\omega_j}}
Substituting eq. \ccc\ and \fff\ in \aaa\ we obtain the relation:
\eqn\appmo{\oint_{b_j}\omega_{PQ}-
\sum_{l=1}^g\overline{T}_{lj} \oint_{a_l}\omega_{PQ}=
2\pi i\overline{\int_Q^P\omega_j}}
On the other side, exploiting similar arguments to evaluate the integral
$\int\omega_{PQ}\wedge\omega_j$ (which necessarily vanishes since $\omega_{PQ}$
is a meromorphic differential and $dz\wedge dz=0$) one has \fay:
\eqn\appmt{\oint_{b_j}\omega_{PQ}-
\sum_{l=1}^gT_{lj} \oint_{a_l}\omega_{PQ}=
2\pi i\int_Q^P\omega_j}
Eqs. \appmo\ and \appmt\ provide a system of linear equations
for the $2g$ unknown periods of $\omega_{PQ}$ around the homology cycles.
After solving it, one finds:
\eqn\aperiods{\oint_{a_k}\omega_{PQ}=
2\pi i \sum_{j=1}^g\left({\rm Im}[T]\right)^{-1}_{jk}{\rm Im}\left[
\int_Q^P\omega_j\right]}
\eqn\bperiods{\oint_{b_j}\omega_{PQ}=
2\pi i \sum_{l=1}^g {\rm Re}[T_{lj}]
\sum_{k=1}^g\left\{\left({\rm Im}[T]\right)^{-1}_{kl}{\rm Im}\left[
\int_Q^P\omega_k\right]\right\}+2\pi i{\rm Re}\left[\int_Q^P\omega_j\right]}
where ${\rm Im}[T]$ is the imaginary part of the period matrix and
$\left({\rm Im}[T]\right)^{-1}_{kl}$ are the elements of its inverse.
From eqs. \aperiods\ and \bperiods\ it is clear that the periods of
$\omega_{PQ}$ are purely imaginary as desired.

Conversely, let us assume that \oconjugprops\ is true.
Evaluating as before $\int_M\omega_{PQ}\wedge\bar\omega_j$
we obtain:
\eqn\ggg{\int_M\omega_{PQ}\wedge\bar\omega_j=
\sum_{l=1}^g\overline{T}_{lj} \oint_{a_l}\omega_{PQ}-
\oint_{b_j}\omega_{PQ}+2\pi i\overline{\int_Q^P\omega_j}}
Moreover, the complex conjugate of eq. \appmt\ becomes with
the help of eqs. \oconjugprops:
\eqn\hhh{\oint_{b_j}\omega_{PQ}=\sum_{l=1}^g
\overline{T}_{jl} \oint_{a_l}\omega_{PQ}+2\pi i
\overline{\int_Q^P\omega_j}}
Substituting the above value of the $b_j-$periods of
$\omega_{PQ}$ in eq. \ggg\ we recover exactly eq.
\newbigleq.

\appendix{B}{}

In this Appendix we show that the surface integral over $\Sigma_g$
of a tensor $f_{z\bar z}(z,y)$ can be rewritten in the form \intexpl.
The material presented here can be found in standard textbooks
\grha -- 
\ref\imgges{
I. M. Gel'fand
and G. E. Shilov, {\it Generalized Functions}, Vol I, Academic Press,
New York and London 1964.
}.
To keep the notations as simple as possible,
we will omit to write explicitly the dependence
of $f_{z\bar z}(z,y)$ on the conjugate variables $\bar z,\bar y$.
We start from
\eqn\intdefapp{I=\int_{\Sigma_g} d^2zf_{z\bar z}(z,y)}
supposing that $f_{z\bar z}(z,y)$ is such that 
the above integral is convergent.
Integrals of the kind \intdefapp\ are to be understood as a sum over all
values of $z\in{\rm\bf CP}^1$ and $y$ for which eq. \curve\
is satisfied, i. e.:
\eqn\gelshi{I=\int_{z\in{\rm\bf CP}^1\atop F(z,y)=0}f_{z\bar z}(z,y)}
Since we assume that our curves are not degenerate, it is possible to write:
\eqn\ggelshi{I=\int_{{\rm\bf CP}^1}d^2z
\int_{{\rm\bf CP}^1}d^2y\sqrt{G}f_{z\bar z}(z,y)\delta^{(2)}(F(z,y))}
where $\sqrt{G}$ is the determinant of the metric in the $y$ domain and the
Dirac $\delta-$function is defined as follows:
\eqn\appbdd{\int_{{\rm\bf CP}^1}d^2y \sqrt{G}f(y)\delta^{(2)}(y-y')=f(y')}
It is now convenient to choose a conformally flat
metric
\eqn\metony{G_{yy}=G_{\bar y \bar y}=0\qquad\qquad
G_{y\bar y}=G_{\bar y y}}
so that eq. \ggelshi\ becomes:
\eqn\gggg{I=\int_{{\rm\bf CP}^1}d^2z
\int_{{\rm\bf CP}^1}d^2yf_{z\bar z}(z,y)\delta^{(2)}_{y\bar y}(F(z,y))}
Locally, we have that:
\eqn\confddf{\delta^{(2)}_{y\bar y}(F(z,y))={1\over 4}
\partial_y\partial_{\bar y}{\rm log}|F(z,y)|^2}
Exploiting the relation
\eqn\fpol{F(z,y)=\prod_{\alpha=0}^{n-1}(y-y^{(\alpha)}(z))}
we see that the zeros of $F(z,y)$ are concentrated
at the branches of $y(z)$, where $F(z,y)$ can be approximated as follows:
\eqn\fapprox{F(z,y)\sim F_y(z,y^{(\alpha)}(z))(y-y^{(\alpha)}(z))+\ldots}
As a consequence:
\eqn\iconc{I=\sum_{\alpha=0}^{n-1}
\int_{{\rm\bf CP}^1}d^2z
\int_{{\rm\bf CP}^1}d^2yf_{z\bar z}(z,y^{(\alpha)}(z))
\delta^{(2)}_{y\bar y}\left(F_y(z,y^{(\alpha)}(z))(y-y^{(\alpha)}(z))\right)}
Performing now the change of variables
\eqn\chvrs{F_y(z,y^{(\alpha)}(z))(y-y^{(\alpha)}(z))=y'}
and integrating over $y'$, we obtain the desired result:
\eqn\finappb{I=\int_{{\rm\bf CP}^1}\sum_{\alpha=0}^{n-1}f_{z\bar z}(
z,y^{(\alpha)}(z))}

\listrefs
\bye